\documentclass[epj]{svjour}
\usepackage{graphics}
\usepackage[latin1]{inputenc}
\usepackage{xcolor}
\usepackage{amssymb}
\usepackage{bm}
\usepackage{enumitem}
\usepackage{citesort}
\usepackage{nicefrac}

\begin{document}
\title{Collinear laser spectroscopy of atomic cadmium}
\subtitle{Extraction of Nuclear Magnetic Dipole and Electric Quadrupole Moments}
\author{Nadja Fr\"ommgen\inst{1} \and Dimiter~L.~Balabanski\inst{2} \and Mark~L.~Bissell\inst{3} \and Jacek Biero\'{n}\inst{4} \and Klaus Blaum \inst{5} \and Bradley Cheal\inst{6} \and Kieran Flanagan\inst{7} \and Stephan Fritzsche\inst{8,9} Christopher Geppert\inst{1,10} \and Michael Hammen\inst{1} \and Magdalena Kowalska \inst{11} \and Kim Kreim\inst{5} \and Andreas Krieger\inst{1,10} \and Rainer Neugart\inst{1,5} \and Gerda Neyens \inst{3} \and Mustafa M.~Rajabali\inst{3} \and Wilfried N\"ortersh\"auser\inst{1,10} \and Jasna Papuga\inst{3} \and Deyan T. Yordanov\inst{5,11,12}}

\institute{Institut f\"ur Kernchemie, Johannes Gutenberg-Universit\"at Mainz, D-55128 Mainz, Germany 
\and ELI-NP, Horia Hulubei National Institute for R\&D in Physics and Nuclear Engineering, 077125 Magurele, Romania
\and Instituut voor Kern- en Stralingsfysica, KU Leuven, Celestijnenlaan 200D, B-3001 Leuven, Belgium
\and Instytut~Fizyki~imienia~Mariana~Smoluchowskiego, Uniwersytet~Jagiello\'nski, ul.~prof.~Stanis\l{}awa~\L{}ojasiewicza~11, 30-348~Krak\'ow,~Poland
\and Max-Planck-Institut f\"ur Kernphysik, Saupfercheckweg 1, D-60117 Heidelberg, Germany
\and Oliver Lodge Laboratory, University of Liverpool, Liverpool, L69 7ZE, UK
\and School of Physics and Astronomy, University of Manchester, Manchester, M13 9PL, UK
\and Helmholtz-Institut Jena, Fr\"obelstieg 3, D-07743 Jena, Germany
\and Theoretisch-Physikalisches Institut, Friedrich-Schiller-Universit\"at Jena, Max-Wien-Platz 1, D-07743 Jena, Germany
\and Institut f\"ur Kernphysik, Technische Universit\"at Darmstadt, D-64289 Darmstadt Germany
\and CERN European Organization for Nuclear Research, Physics Department, CH-1211 Geneva 23, Switzerland
\and Institut de Physique Nucl\'eaire, Orsay, IN2P3/CNRS, 91406 Orsay Cedex, France 
}
\date{Received: date / Revised version: date}
%
\abstract{Hyperfine structure $A$ and $B$ factors of the atomic $5s\,5p\,\; ^3\rm{P}_2 \rightarrow 5s\,6s\,\; ^3\rm{S}_1$ transition are determined from collinear laser spectroscopy data of $^{107-123}$Cd and $^{111m-123m}$Cd. Nuclear magnetic moments and electric quadrupole moments are extracted using reference dipole moments and calculated electric field gradients, respectively. The hyperfine structure anomaly for isotopes with $s_{\nicefrac{1}{2}}$ and $d_{\nicefrac{5}{2}}$ nuclear ground states and isomeric $h_{\nicefrac{11}{2}}$ states is evaluated and a linear relationship is observed for all nuclear states except $s_{\nicefrac{1}{2}}$. This corresponds to the Moskowitz-Lombardi rule that was established in the mercury region of the nuclear chart but in the case of cadmium the slope is distinctively smaller than for mercury. In total four atomic and ionic levels were analyzed and all of them exhibit a similar behaviour. The electric field gradient for the atomic $5s\,5p\,\; ^3\mathrm{P}_2$ level is derived from multi-configuration Dirac-Hartree-Fock calculations in order to evaluate the spectroscopic nuclear quadrupole moments. The results are consistent with those obtained in an ionic transition and based on a similar calculation.  
} 
\maketitle

\section{Introduction}
\label{intro}
Atomic spectroscopy of short-lived isotopes is a unique tool to investigate nuclear moments, spins and charge 
radii in a nuclear-model independent way. The magnetic moments provide access to the single-particle structure of nuclei while information on the nuclear shape and collectivity is imprinted in the quadru\-pole moments. Combined with the spin, extracted from the hyperfine structure, these nuclear properties and their evolution along an isotopic chain provide invaluable insight into the forces governing the nucleus. Experimental accuracies of the hyperfine structure (hfs) splittings are 
often at the $10^{-4}$ level if high-resolution laser spectroscopy is applied. Extraction of the nuclear magnetic
 dipole moment $\mu$ and the spectroscopic electric quadrupole moment $Q_{\rm s}$ directly from the hyperfine $A$
 and $B$ parameters of a particular isotope requires a very good knowledge of the hyperfine fields, i.e. the magnetic field $B_e(0)$ and the electric field gradient $V_{zz}(0)$, induced by the electrons at the site of the nucleus. Alternatively, the moment of the isotope $X$ can be calculated if the respective nuclear moment is known for at least one reference isotope of the same isotopic chain. In the case of the nuclear magnetic moment, the corresponding relation 
\begin{equation}
  \mu_X = \frac{A_X}{A_{\rm Ref}} \frac{I_X}{I_{\rm Ref}} \mu_{\rm Ref} ,
	\label{eq:magnMomentARatio}
\end{equation}
can be used to determine the magnetic moment from the ratio of the magnetic coupling constants $A_X$ and $A_{\rm Ref}$, measured by laser spectroscopy, and the corresponding nuclear spins $I_X$ and $I_{\rm Ref}$. Magnetic dipole moments of stable and reasonably long-lived isotopes are usually known with high accuracy from nuclear magnetic resonance (NMR) or atomic beam magnetic resonance experiments. 
However, Eq.\,(\ref{eq:magnMomentARatio}) is based on a point-dipole approximation for the nuclear magnetic moment. In reality the magnetic moment is distributed over the finite size of the nucleus. The influence of this spatial distribution is a deviation of the measured dipole coupling constant $A$ from that of a nucleus with identical charge distribution but point-like dipole moment, $A_{\mathrm{p}}$. This is taken into account by the Bohr-Weisskopf correction $\varepsilon_{\rm BW}$ according to \cite{BohrWeisskopf1950}
\begin{equation}
  A = A_{\mathrm{p}} \cdot \left( 1 + \varepsilon_{\rm BW} \right).
\end{equation}
The correction coefficient $\varepsilon_{\rm BW}$ is isotope dependent and reflects the relative contributions of spin and orbital angular momentum of the active valence nucleons to the magnetic moment. As point nuclei do not exist and theoretical calculations of $A$ factors are not accurate enough, the ``Bohr-Weisskopf effect'' is only seen in small deviations of the $A$-factor ratios between different isotopes from the ratios given by the magnetic moments. 
These deviations
\begin{equation}
{^{1}{\rm \Delta}^{2}} \approx \varepsilon_{\rm BW}^{\rm 1} - \varepsilon_{\rm BW}^{2}
\label{eq:hfa4}
\end{equation}
are known as hyperfine structure anomalies 
and Eq.\,(\ref{eq:magnMomentARatio}) has to be modified to
\begin{equation}
\mu_X = \frac{A_X}{A_{\rm Ref}} \frac{I_X}{I_{\rm Ref}} \; \mu_{\rm Ref} \; \left( 1 + {^{\rm Ref}{\rm \Delta}^{X}} \right).
			\label{eq:hfa3}
\end{equation}
The hfs anomaly is typically of the order of $10^{-4}$ for light elements to $10^{-2}$ for heavy elements and is usually tabulated in percent.

The study of hfs anomalies requires very accurate and independent measurements of magnetic moments and hfs parameters. This is the reason why systematic data are available 
only in a few cases. One of the prominent examples is the series of mercury isotopes between $^{193}$Hg and $^{203}$Hg for which Moskowitz and Lombardi \cite{Mos73} realized that the hfs anomaly follows a surprisingly simple rule. Plotted against the difference of the inverse magnetic moments, the anomalies
\begin{equation}
{^{1}{\rm \Delta}^{2}} = \frac{A_1}{A_2}\,\frac{I_1}{I_2}\,\frac{\mu_2}{\mu_1} - 1
   \label{eq:hfsAnom}
\end{equation}
reveal a linear dependence corresponding to the relationship 
\begin{equation}
{^1\mathrm{\Delta}^2}=\alpha \left( \frac{1}{\mu_1} - \frac{1}{\mu_2} \right),
   \label{eq:LinRel}
\end{equation}
which suggests that for the individual isotopes
\begin{equation}
   \varepsilon_{\rm BW} = \frac{\alpha}{\mu_I}\,. 
	 \label{eq:ML-Rule}
\end{equation}
For the mercury $6s\,6p\;\, ^3\mathrm{P}_1$ level the proportionality constant was found to be $\alpha = \pm 1.0 \cdot 10^{-2} \mu_N$, depending on the coupling $I=\ell \pm \nicefrac{1}{2}$, respectively, of orbital and spin angular momentum in the shell-model description of the odd-neutron state. 
According to the rule (\ref{eq:ML-Rule}) nuclear states in the tin region like $d_{\nicefrac{5}{2}},\, h_{\nicefrac{11}{2}}$, resulting from the coupling $I=\ell+\nicefrac{1}{2}$, and $d_{\nicefrac{3}{2}}$ from $I=\ell-\nicefrac{1}{2}$ should also exhibit equal absolute values of $\alpha$ with opposite sign. 

While a similar behavior was observed for a few isotopes of odd-$Z$ elements in the mercury region \cite{Mos82}, namely $^{191,193}$Ir, $^{196-199}$Au, $^{203,205}$Tl, no conclusive picture about the validity of the rule has emerged from scattered data on elements in other regions of the nuclear chart \cite{Per05}. It seems desirable to obtain more experimental information and to better understand the nuclear model assumptions to be made for a theoretical validation of relations (\ref{eq:LinRel}) and (\ref{eq:ML-Rule}). Nevertheless, the Moskowitz-Lombardi rule has often been used to estimate uncertainties arising from hyperfine anomalies for nuclear moments extracted from hyperfine splittings (see, e.g., \cite{Eks80,Mue83}). 

The cadmium chain is in many respects similar to the mercury case. Cd ($Z=48$) is one proton pair below the $Z=50$ shell closure, as is Hg ($Z=80$) below the next shell closure at $Z=82$. In both regions isomeric states exist due to the opposite-parity high-spin orbitals, $h_{\nicefrac{11}{2}}$ below $N=82$ and $i_{\nicefrac{13}{2}}$ below $N=126$. Moreover, accurate measurements of nuclear moments exist for $^{107,109,111,113}$Cd \cite{Cha69}, and for $^{115}$Cd as well as for the isomers $^{113m,115m}$Cd \cite{Spe72}, covering nuclear spins of $I=\nicefrac{5}{2}$, \nicefrac{1}{2} and \nicefrac{11}{2}. This facilitates an analysis of hfs anomalies.  

While magnetic moments are usually extracted from measured hfs $A$ factors according to Eq.\,(\ref{eq:magnMomentARatio}) by using experimental reference values of both quantities for one isotope, a similar approach is not possible for the nuclear quadrupole moments since directly measured reference values of $Q_{\rm s}$ are not available. Instead, the extraction of $Q_{\rm s}$ from atomic $B$ factors has to rely on calculations of the electric field gradient $V_{zz}$ at the site of the nucleus. 
Traditionally such calculations have been based on classical hfs theory, using semi-empirical expectation values $\langle r^{-3} \rangle$ for a particular valence electron configuration. To date, all tabulated quadrupole moments of cadmium isotopes are taken from the early work of McDermott et al., in particular \cite{Lau69}. The field gradient was calculated from the dipole and quadrupole interaction constants of the $p_{\nicefrac{3}{2}}$ electron in the hfs of the $^3P_2$ and $^3P_1$ states of $^{109}$Cd, and a typical 10\% uncertainty was estimated, covering the range of values resulting from alternative evaluations of $\langle r^{-3} \rangle$. Sternheimer-type shielding corrections were neglected. It was a goal of the present work to provide more reliable electric field gradients of different atomic states from modern hfs theory and thus obtain a reference for the evaluation of a new set of Cd quadrupole moments.

Already in Ref. \cite{Yor13} the quadrupole moments were based on a $V_{zz}$ value taken from Dirac-Hartree-Fock calculations in the Cd$^+$ (Cd II) $5p\,\;^2\mathrm{P}_{\nicefrac{3}{2}}$ level. In Sect.\,\ref{sec:efg} we give details of these calculations and of similar ones for the atomic (Cd I) $5s\,5p\;\, ^3\mathrm{P}_2$ level with a comparison of the accuracy and consistency of the results. For reasons discussed in Sect.\,\ref{moments}, we rely on the values calculated for Cd II and the results given here are consistent with those already published in \cite{Yor13}.

Here we report on laser spectroscopy measurements in the atomic $5s\,5p\;\, ^3\mathrm{P}_2 \rightarrow 5s\,6s\;\, ^3\mathrm{S}_1$ transition for $^{106-124}$Cd and $^{126}$Cd. 
Nuclear ground states as well as long-lived isomeric states were investigated. The focus of this paper is on the extraction of the nuclear parameters from the atomic spectra. Magnetic moments are extracted from hfs $A$ factors using Eq.\,(\ref{eq:magnMomentARatio}) and this is combined with an analysis of the hfs anomaly and an examination of the applicability of the Moskowitz-Lombardi rule. Quadrupole moments are extracted from $B$ factors using newly calculated electric field gradients obtained from multi-configuration Dirac-Hartree-Fock theory. The reliability of these calculations is verified by comparing the results for the atomic level investigated here with those for the level in the Cd$^+$ ion reported in \cite{Yor13}.

\section{Experimental setup}
\label{sec:exp}
The measurements were performed at the on-line isotope separator ISOLDE at CERN. The cadmium isotopes were produced by 1.4-GeV protons impinging on a uranium carbide target and ionized by resonance laser ionization in the laser ion source of ISOLDE. After acceleration to 50~keV the ions were mass separated in the General Purpose Separator (GPS) and guided to the collinear laser spectroscopy setup (COLLAPS). 
At the COLLAPS beam line, the fast ion beam is collinearly superimposed with the laser beam and subsequently the ions are neutralized by charge-exchange reactions with sodium atoms in a vapour cell to which the post-acceleration voltage was applied. Resonance fluorescence light was observed in the fluorescence detection region (FDR), mounted adjacently to the exchange cell, by four photomultiplier tubes. The layout of the FDR system is described in \cite{Kre14}.

Spectroscopy was performed in the atomic (Cd I) transition from the meta\-stable $5s\,5p\,\; ^3\rm{P}_2$ level, partially populated in the charge-exchange process, to the excited $5s\,6s$ $^3\rm{S}_1$ level. This corresponds to a transition wave number of 19657.031~$\rm{cm}^{-1}$ \cite{Burns1956}. The required laser light at a wavelength of 508.7\,nm was produced by a dye laser, operated with Coumarin 521, which was pumped by an argon-ion laser operated at 488\,nm. The laser frequency was constantly measured with a wave\-meter and locked to an external cavity which was again stabilised in length to a frequency stabilised helium:neon laser. Thus a stability of the dye laser of the order of a few MHz over several hours was obtained. A laser power of about 1~mW was used for spectroscopy.

While the laser was fixed at a frequency $\nu_{\rm{L}}$ in the laboratory system, the velocity of the ions was manipulated by applying an additional tunable post-acceleration voltage in the range of $\pm 10$\,keV to the charge-exchange cell. Thus the laser frequency is shifted in the rest frame of the ion according to the Doppler formula to $\nu_c = \nu_{\rm{L}} \gamma (1-\beta)$.
The velocity $\beta = \upsilon / c$ in terms of the speed of light is obtained from the difference of the (positive) acceleration potential at the ISOLDE ion source and the post-acceleration voltage at the charge-exchange cell. The time dilation factor is calculated according to $\gamma=(1-\beta^2)^{-\nicefrac{1}{2}}$. In order to reach sufficient accuracy, the calibration reported in \cite{Kri11} is applied for the ISOLDE high voltage and the post-acceleration voltage is measured repeatedly with a Julie Research 10-kV high-voltage divider providing a relative accuracy of  $1.2 \cdot 10^{-4}$. 

\section{Analysis and theory}
\label{sec:ana}
\begin{figure}
\resizebox{0.48\textwidth}{!}{%
\includegraphics{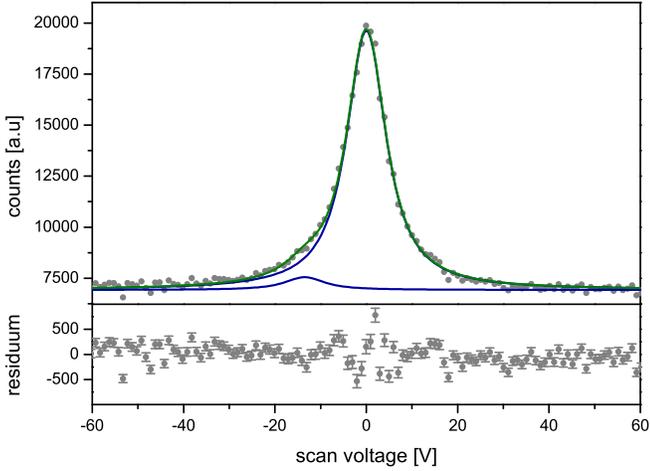}
}
\caption{Spectrum of the reference isotope $^{114}$Cd in the $5s\,5p\,\; ^3\rm{P}_2 \rightarrow 5s\,6s\,\; ^3\rm{S}_1$ transition. The fitted curve is the sum of two Voigt profiles -- also shown individually -- with a relative position of 13.5~V to each other. The lower panel shows the residuum between the experimental data and the fit. Typical total linewidth (Lorentzian + Gaussian contribution) is about 30-40\,MHz.
 }
\label{fig:line profile}
\end{figure}

Sufficient production rates of $>10^5$\,ions/s to perform laser spectroscopy measurements on atomic Cd with continuous ion beams were reached for $^{106-124,126}$Cd. In addition to the nuclear ground states the $I=\nicefrac{11}{2}^-$ isomers were observed for all odd-mass isotopes from $^{111}$Cd upwards. 
The isotope with the highest natural abundance in the cadmium chain, $^{114}$Cd, was chosen as the reference for isotope shift measurements. The spectra of the other isotopes were recorded in alternation with $^{114}$Cd. A typical spectrum of the reference isotope is shown in Fig.~\ref{fig:line profile} as a function of the post-acceleration voltage at the charge-exchange cell. It exhibits an asymmetric tail at lower voltages, i.e. higher beam energies. 
Thus it is produced by ions in the beam that need additional post-acceleration to be in resonance because they suffer a loss of energy in atomic collisions before they interact with the laser beam. The spectrum can be well fitted by two Voigt profiles with identical line shape parameters. The complete set of reference spectra were first analyzed using these Voigt profiles with free distances and relative intensities. It was observed that the distance between the two peaks was approximately constant with an average value of $\Delta U= 13.5$\,V but the common width and relative intensity changed slightly during the four days of beam time. Such a tail is often observed in collinear laser spectroscopy and usually ascribed to energy loss in the charge-exchange reaction or ion-atom collisions along the beam line. However, in this case the uncommonly large separation is not explainable by a single process. Instead, multiple side peaks shifted in the energy by integer multiples of the transition energy would give a similarly good description of the line shape. Hence, the single side peak at this large distance can probably be understood as a feature effectively mimicking the resulting tail from multiple collisional processes. 
To avoid artificial shifts of the line center in the fitting routine, all spectra -- including those of the reference isotope -- were finally analyzed with the double Voigt profile of common linewidth and a fixed peak distance.
The reference isotope was fitted with free intensity ratios. The spectra of all other isotopes were analyzed twice: once with linewidth parameters and relative intensities as obtained from the preceding reference scan and once with parameters as determined in the succeeding reference scan. The resonance positions and hyperfine parameters obtained in the two fits were averaged and the larger of the two fitting uncertainties was adopted. 
 
\begin{figure*}
\resizebox{1.0\textwidth}{!}{%
\includegraphics{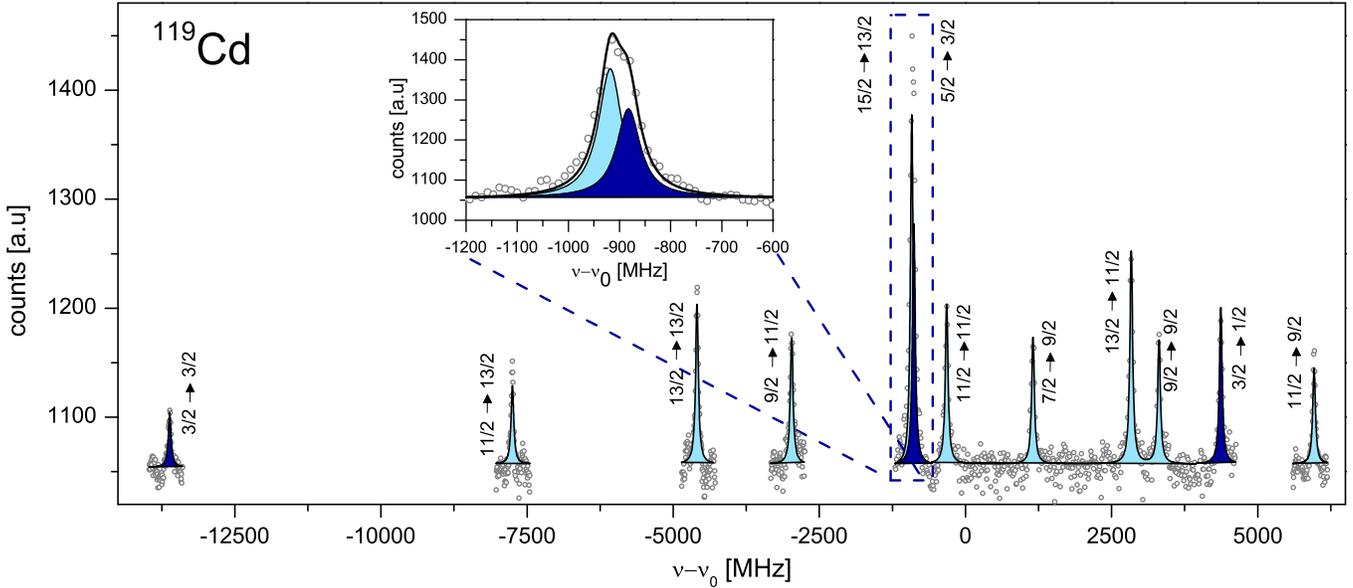}
}
\caption{(color online) Hyperfine structure of the nuclear ground state (dark blue) and the isomeric $\nicefrac{11}{2}^-$state (light blue) of $^{119}$Cd in the $5s\,5p\,\; ^3\rm{P}_2 \rightarrow 5s\,6s\,\; ^3\rm{S}_1$ transition. Both states can be clearly distinguished from each other. The number of hyperfine lines arising from the nuclear ground state unambiguously yields a nuclear spin of $I=\nicefrac{1}{2}$.}
\label{fig:119cd}
\end{figure*}

The recorded hyperfine spectra of the odd isotopes were fitted using the centre of gravity (cg) and the $A$ and $B$ factors of both electronic levels as free parameters. 
The  hfs shift of each sublevel with total angular momentum $\vec{F}=\vec{I}+\vec{J}$ composed of electronic angular momentum $J$ and nuclear spin $I$ and corresponding factor $C=F(F+1)-I(I+1)-J(J+1)$ was calculated according to the first order hyperfine energy 
\begin{equation}
   \Delta E_{\rm{hfs}}=A\frac{C}{2}+B\cdot\frac{\frac{3}{4}C(C+1)-I(I+1)J(J+1)}{2I(2I-1)J(2J-1)} \ . 
	\label{eq:hfsForm}
\end{equation}
In the fit function, these shifts determine the spectral line positions relative to the centre of gravity and each hyperfine component is again composed of the main peak and the tailing peak as discussed above.\\
The hyperfine parameters $A$ and $B$ of the upper and the lower electronic levels were fitted independently for all isotopes except for $^{119}$Cd for which the spectrum is shown in Fig.~\ref{fig:119cd}. In this isotope, due to overlapping resonances of the isomer and the nuclear ground state, shown and enlarged in Fig.~\ref{fig:119cd}, the ratio of the two $A$ factors was fixed to the error-weighted mean value of the other isotopes as discussed in Sect.\,\ref{sec:hfs}. However, in an independent analysis, the spectrum could also be analyzed without this constraint by fitting both nuclear states simultaneously and both results were in agreement within uncertainties.

The uncertainty of the acceleration voltage gives rise to systematic uncertainties of the determined isotope shifts. Compared to these, the contributions from atomic masses and the laser frequency are negligible.  

\section{Results and discussion}
\label{sec:results}
The fitting procedure yields the hyperfine parameters $A$ and $B$ and the centroid transition frequencies $\nu_{\rm{cg}}$ relative to $^{114}$Cd. From these parameters, the nuclear observables can be extracted if the hyperfine fields are known at the site of the nucleus. This section describes how these atomic parameters were determined.

\subsection{Nuclear spin}
\label{sec:spin}
The nuclear spin of an isotope can be directly deduced from the number of peaks 
if $I < J$ (in the case $I=\nicefrac{1}{2}$ also for $I=J$). In the $5s\,5p\, ^3\mathrm{P}_2\,$ $(J=2) \,\rightarrow\, 5s\,6s\;\, ^3\mathrm{S}_1 (J'=1)$ transition, for example, this determination of the nuclear spin is possible for $I=\nicefrac{1}{2}$ and $I=\nicefrac{3}{2}$. In these cases three peaks and eight peaks are observed, respectively (the case $I=1$ cannot occur for the Cd isotopes). For larger spins nine hyperfine components appear. The situation is illustrated for the ground state of $^{119}$Cd (dark blue) and its $\nicefrac{11}{2}^-$ isomer (light blue) in Fig.~\ref{fig:119cd}. Both these states can be clearly distinguished from each other. Nine peaks belong to the $\nicefrac{11}{2}^-$ isomer, exhibiting the same hyperfine pattern as the other isomers. Only three peaks remain for the ground state of $^{119}$Cd which therefore must have a nuclear spin of $I=\nicefrac{1}{2}$. This clarifies a misassignment in literature where $^{119}$Cd was assigned a spin of $I=\nicefrac{3}{2}$ based on $\gamma$-ray studies \cite{Sym09}. This is consistent with conclusions drawn from the measurements on Cd$^+$ ions published previously \cite{Yor13}.
For all other isotopes, the formerly (partly tentatively) reported nuclear spins were confirmed by the  hfs measurements. The nuclear spins are listed in Table~\ref{tab:moments}. Revisions of spin assignment occur at $^{111}$Cd from $I=\nicefrac{5}{2}$ to $I=\nicefrac{1}{2}$ and at $^{121}$Cd from $I=\nicefrac{1}{2}$ to $I=\nicefrac{3}{2}$ for the odd isotopes. All isomers have the spin $I=\nicefrac{11}{2}$. According to the spins, the unpaired neutron should occupy the $2d_{\nicefrac{5}{2}}$, $3s_{\nicefrac{1}{2}}$, $2d_{\nicefrac{3}{2}}$ and $1h_{\nicefrac{11}{2}}$ shell-model orbitals, respectively.

\subsection{Hyperfine structure parameters}
\label{sec:hfs}
The extracted  hfs parameters $A$ and $B$ of the studied transition are presented in Table~\ref{tab:hfs factors}. The quoted uncertainties include the statistical and the systematic uncertainties added in quadrature. The latter are caused by the uncertainty in the determination of the post-acceleration voltage applied to the charge-exchange cell. The quadrupole interaction for a pure $5s6s\, ^3\rm{S}_{1}$ state should be zero. However, since the total angular momentum is $J=1$, configuration mixing with other $J=1$ states results in a (usually small) finite $B$ value. Therefore, the $B$ parameter for the $5s6s\, ^3\rm{S}_{1}$ electronic level was not fixed to be zero in the fitting routine. The resulting $B$ factors of this level are nevertheless very small, have uncertainties much larger than their magnitudes and are thus not included in Table~\ref{tab:hfs factors}.

In Fig.~\ref{fig:A ratio} the ratio of the $A$ factors $A(5s\,5p\,\; ^3\rm{P}_2)$ and $A(5s\,6s\,\; ^3\rm{S}_1)$ is plotted. The values of the ground and isomeric state of $^{119}$Cd are missing since in these cases the ratio was constrained in the fitting as discussed above. 
The ratio is constant within uncertainties, showing no appreciable amount of isotope or spin dependence -- a fact that supports the choice of an average $A$-factor ratio for fitting the spectrum of $^{119,119m}$Cd. The error-weighted mean $A$ factor ratio of all isotopes is 0.42479(3), indicated by a dashed line in Fig.~\ref{fig:A ratio}.

Other results available from literature are included in Table~\ref{tab:hfs factors}. They are all in agreement with the values obtained in this work. While $A$ factors for the $5s\,5p\,\; ^3\rm{P}_2$ level in $^{111}$Cd and $^{113}$Cd (both $I=\nicefrac{1}{2}$) are available in the literature with orders of magnitude higher precision \cite{Fau60}, obtained with the atomic beam magnetic resonance technique \cite{Fau60}, corresponding values for the other isotopes are from pres\-sure-scanned Fabry-Pérot spectroscopy \cite{Bri76} and from collinear laser spectroscopy performed at COLLAPS in 1989 \cite{Boo89}. Both have larger or comparable uncertainties than the values reported here. Accurate NMR values for the nuclear magnetic moment are available for several isotopes and are included in Table~\ref{tab:hfs factors} since they are used in the following to analyse hfs anomalies and serve as a reference for the extraction of nuclear moments.
\begin{table*}
\centering
\caption{$A$ and $B$ factors of the hfs splitting of the $5s\,5p\,\; ^3\rm{P}_2$ and the $5s\,6s\,\; ^3\rm{S}_1$ electronic levels in the cadmium atom for isotopes with mass number $A=Z+N$, half-life $T_{\nicefrac{1}{2}}$ and spin and parity $I^\pi$. The $B$ factor of the upper electronic level is within its uncertainty compatible with zero and is not included in the table. The listed total uncertainties are the quadratic sum of the statistical uncertainty and the systematic uncertainty as discussed in the text. For $^{119}$Cd and $^{119\rm{m}}$Cd the ratio of the $A$ factors was constrained to the average of all other isotopes. Literature values are listed where available with their references (square brackets) for comparison.}
\label{tab:hfs factors}      
\begin{tabular}{c c c r@{}r@{}l r@{}r@{}l r@{}r@{}l r@{}r}
\hline\noalign{\smallskip}
 & & & \multicolumn{3}{c}{$5s\,5p\,\; ^3\rm{P}_2$} & \multicolumn{3}{c}{$5s\,6s\,\; ^3\rm{S}_1$} & \multicolumn{3}{c}{$5s\,5p\,\; ^3\rm{P}_2$} & \multicolumn{2}{c}{$\mu_{\rm NMR}$ \cite{Sto05}}\\
\noalign{\smallskip} 
$Z+N$ & $T_{\nicefrac{1}{2}}$ & $I^{\pi}$ & \multicolumn{3}{c}{$A$ (MHz)} & \multicolumn{3}{c}{$A$ (MHz)} & \multicolumn{3}{c}{$B$ (MHz)} &  \multicolumn{2}{c}{$(\mu_{\rm N})$} \\
\noalign{\smallskip}\hline \hline\noalign{\smallskip}
107 & 6.50\,h &$\nicefrac{5}{2}^+$ & $-682.0$     & (4) &              & $-1605.6$ & (7) &           & $+290$ & (3) && $-0.6150554$ & (11)\\
109 & 461.4 d & $\nicefrac{5}{2}^+$ & $-917.5$     & (1) &              & $-2159.9$ & (4) &           & $+296$ & (2) && $-0.8278461$ & (15)\\
111 & stable & $\nicefrac{1}{2}^+$ & $-3292.7$    & (5) &              & $-7750.9$ & (10)&           &        &     && $-0.5948861$ & (8) \\
    & &                      & $-3292.9364$ & (2) &$^{\rm a}$    & $-7750.5$ & (18)& $^{\rm b}$&        &     &&              & \\
113 & stable & $\nicefrac{1}{2}^+$ & $-3444.5$    & (6) &              & $-8108.8$ & (11)&           &        &     && $-0.6223009$ & (9)\\
    &   &                  & $-3444.6344$ & (16)& $^{\rm a}$   &           &     &           &        &     &&              &\\
115 &  53.46\,h & $\nicefrac{1}{2}^+$ & $-3588.6$    & (8) &              & $-8444.3$ & (28)&           &        &     && $-0.6484259$ & (12)\\
117 & 2.49\,h & $\nicefrac{1}{2}^+$ & $-4116.7$    & (8) &              & $-9687.9$ & (17)&           & & &&&\\
    &    &                 & $-4116.1$    &(63) &$^{\rm c}$    & $-9698$   & (10)&$^{\rm c}$ & & &&&\\
119 & 2.69\,min &$\nicefrac{1}{2}^+$ & $-5091.8$    &(12) &              & & & & &&&&\\
121 & 13.5\,s & $\nicefrac{3}{2}^+$ & $+1156.4$    & (4) &              & $+2722.0$ & (6) &           & $-141$ & (7) &&&\\
123 & 2.1\,s & $\nicefrac{3}{2}^+$ & $+1457.2$    & (8) &              & $+3431.9$ & (13)&           & $+31$  & (4) &&&\\
\noalign{\smallskip}\hline\noalign{\smallskip}
111m & 48.50\,min & $\nicefrac{11}{2}^-$ & $-556.5$ & (1) &                & $-1310.3$ & (2) &           & $-363$ & (2) && $-1.1051$    & (4)\\
113m & 14.1\,a & $\nicefrac{11}{2}^-$ & $-547.8$ & (1) &                & $-1289.5$ & (2) &           & $-299$ & (2) && $-1.087784$ & (2)\\
115m & 44.56\,d & $\nicefrac{11}{2}^-$ & $-524.3$ & (2) &                & $-1234.2$ & (2) &           & $-234$ & (2) && $-1.0410343$ & (15)\\
117m & 3.36\,h & $\nicefrac{11}{2}^-$ & $-502.0$ & (1) &                & $-1181.8$ & (3) &           & $-156$ & (1) & &\\
     & &                     & $-501.4$ & (6) &$^{\rm c}$      & $-1181.5$ & (8) &$^{\rm c}$ & $-144$ & (15)&$^{\rm c}$ &&\\
119m & 2.20\,min & $\nicefrac{11}{2}^-$ & $-485.3$ & (2) &                &           &     &           & $-67$  & (1) &&&\\
121m & 8.3\,s & $\nicefrac{11}{2}^-$ & $-508.5$ & (2) &                & $-1197.2$ & (3) &           & $+3$   & (3) &&&\\
123m & 1.82\,s & $\nicefrac{11}{2}^-$ & $-504.0$ & (3) &                & $-1186.5$ & (5) &           & $+63$  & (6) &&&\\
\noalign{\smallskip}\hline
\multicolumn{13}{l}{$^{\rm a}$ Faust \textit{et al.} \cite{Fau60}}\\
\multicolumn{13}{l}{$^{\rm b}$ Brimicombe \textit{et al.} \cite{Bri76}}\\
\multicolumn{13}{l}{$^{\rm c}$ Boos \cite{Boo89}}\\
\end{tabular}
\end{table*}

\definecolor{darkgreen}{rgb}{0,.5,0}
\begin{figure}
\resizebox{0.48\textwidth}{!}{%
\includegraphics{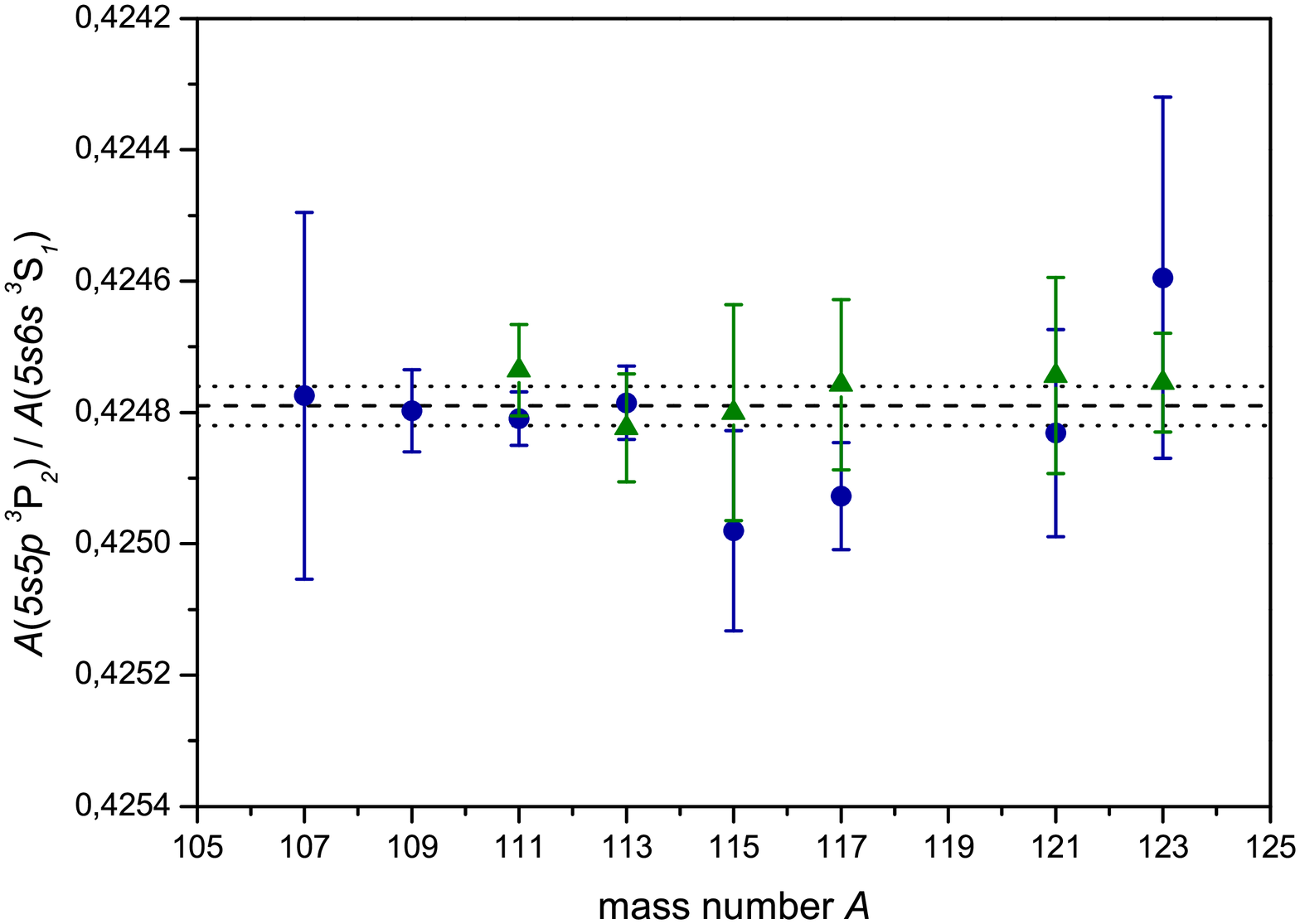}
}
\caption{Ratio of the $A$ factors of the $5s\,5p\,\; ^3\rm{P}_2$ and the $5s\,6s\,\; ^3\rm{S}_1$ electronic levels obtained in the nuclear ground state ($\color{blue}\bullet$) and the isomeric ($\color{darkgreen}\blacktriangle$) state. The dashed line indicates the error weighted mean value and the dotted lines represent the $1\sigma$ uncertainty.}
\label{fig:A ratio}
\end{figure}

\subsection{Hyperfine structure anomaly and the Moskowitz-Lombardi rule}
\label{sec:hfa}
A change of the magnetization distribution from one isotope to the other gives rise to an (isotopic) hfs anomaly $^{1}{\rm \Delta}^{2}$ as defined in Eq.\,(\ref{eq:hfsAnom}).
For the isotopes with precisely known nuclear magnetic moments obtained by NMR, the (isotopic) hfs ano\-malies $^{111}{\rm \Delta}^{M}$ for the isotopes with mass number\footnote{Here we use $M$ instead of the usually used letter $A$ for the mass number of an isotope to avoid confusion with the hyperfine $A$ factor} $M$ with respect to $^{111}$Cd are calculated from the fitted  hfs parameters of the initial and the final electronic level $A(^3{\rm P}_2)$ and  $A(^3{\rm S}_1)$, respectively. These anomalies are listed in Table~\ref{tab:hyperfineanomaly} and are compared to values from the literature for the $5s\,5p \,\; ^3{\rm P}_1$ level. 

Even though the ratio of the $A$ factors in the upper and the lower level of the transtion as plotted in Fig.\,\ref{fig:A ratio} is constant within the measurement uncertainty, the individual levels exhibit a substantial amount of hfs anomaly. This is however masked by the fact that the so-called differential hfs anomaly, being the difference in the hfs anomaly between the initial and the final level of the atomic transition, is very small. This leads to the conclusion that the hfs anomaly arises dominantly from the $5s$ electron which is common to both atomic levels. The $5p$ electron in the $^3\rm{P}_2$ level must have largely $5p\,_{\nicefrac{3}{2}}$ character to build up the $J=2$ level and thus has a probability density vanishing in the nuclear interior. Consequently, also the $6s$ electron cannot have a considerable contribution, otherwise the differential hfs anomaly between the two electronic levels would be larger. A comparison between the observed isotopic hfs anomaly in the $5s\,5p\,\; ^3\rm{P}_2$ level and the one in the $5s\,5p\,\; ^3\rm{P}_1$ fine structure level (lower rows of Table~\ref{tab:hyperfineanomaly}) indicates that the hfs anomaly in the $^3\rm{P}_1$ level is even slightly smaller. Since this fine structure level must have some $p\,_{\nicefrac{1}{2}}$ character, the $p\,_{\nicefrac{1}{2}}$ contribution must be responsible for the difference and seems to compensate partially the hfs anomaly caused by the $5s\,_{\nicefrac{1}{2}}$ electron.

This line of reasoning can be supported by relating the $A$ factors in the two-electron systems of the cadmium atom to the single-electron $a$-factor in the ion. This can be performed 
assuming Russel-Saunders (LS) coupling and neglecting contributions from the $5p$ electron. This leads to \cite{Kop58} 
\begin{eqnarray}
A\left(5s\,6s\, ^3\mathrm{S}_1\right) \approx \frac{1}{2}  \left( a_{5s\,} + a_{6s} \right) \label{eq:TwoElectronA3S} \\
A\left(5s\,5p\, ^3\mathrm{P}_2\right) \approx a_{5s\,} \frac{1}{2(\ell+1)}= \frac{a_{5s\,}}{4}. \label{eq:TwoElectronA3P}
\end{eqnarray}
Using the $A$ and $g$ factors of the ion \cite{Yor13} and the atomic levels observed here, we find 
\begin{eqnarray}
\label{eq:TwoElectronA}
A/g \left(5s\,\, ^2\mathrm{S}_{\nicefrac{1}{2}}\right) \approx 12,200\,\mathrm{MHz},             \label{eq:AgExpIon}\\
A/g \left(5s\,6s\, ^3\mathrm{S}_{1}\right) \approx 6,500\,\mathrm{MHz}, \,\mathrm{and} \label{eq:AgExpAtomS}\\ 
A/g \left(5s\,5p\, ^3\mathrm{P}_{2}\right) \approx 2,800\,\mathrm{MHz}.              \label{eq:AgExpAtomP}
\end{eqnarray}
Based on the first value and the above quoted relation (\ref{eq:TwoElectronA3P}) one would expect an $A/g$-value for the $5s\,5p\,\; ^3\mathrm{P}_{2}$ level of $\approx 3,050$\,MHz which is even 10\% larger than the experimental value, indicating that indeed the hyperfine interaction arises almost exclusively from the $5s$ electron, reduced  by a weak shielding from the $5p$ electron. Using Eq.\,(\ref{eq:TwoElectronA3S}), it follows that the $6s$ electron has a contribution that is of the order of 10\% of the $5s$ electron.

The isotopic hfs anomaly for the isotope pairs $^{111,113}$Cd and $^{111,115}$Cd is close to zero in all electronic levels. This is expected since these isotopes have identical nuclear spin $I=\nicefrac{1}{2}$ and should therefore have a similar magnetization distribution. For the isotopes with a larger nuclear spin, larger values are observed but again isotopes having identical nuclear spin and parity, e.g. $^{107,109}$Cd ($\nicefrac{5}{2}^+$) or the $\nicefrac{11}{2}^-$ isomers, exhibit $^{111}{\rm \Delta}^{M}$ values that are similar in size.

\begin{table}
\centering
\caption{Isotopic hyperfine anomalies $^{111}{\rm \Delta}^{M}$ of isotopes with mass number $M$ and spin $I$ relative to $^{111}$Cd in the electronic levels $5s\,5p\,\; ^3\textrm{P}_2$ and $5s\,6s\,\; ^3\textrm{S}_1$. Listed are the isotopes for which high-precision values of the nuclear magnetic moment are available from literature. The hyperfine parameters $A$ obtained in this work, which were used for the calculations according to Eq.\,(\ref{eq:hfsAnom}), are listed in Table~\ref{tab:hfs factors}. Literature values for the $5s\,5p\,\; ^3\textrm{P}_1$ level are included for discussion.}
\label{tab:hyperfineanomaly}
\begin{tabular}{c r r r c }
\hline\noalign{\smallskip}
Isotope & &  $5s\,5p\,\; ^3\textrm{P}_2$ & $5s\,6s\,\; ^3\textrm{S}_1$ & \\
\noalign{\smallskip} 
$M$ & $I^\pi$ & $^{111}{\rm \Delta}^{M}~(\%)$ &  $^{111}{\rm \Delta}^{M}~\left[\%\right]$ & Ref. \\
\noalign{\smallskip}\hline \hline\noalign{\smallskip}
107 & $\nicefrac{5}{2}^+$ & -0.17(5) & -0.18(4) & this work \\
109 & $\nicefrac{5}{2}^+$ & -0.12(1) & -0.12(1) & this work \\
113 & $\nicefrac{1}{2}^+$ & 0.00(1) & -0.01(1)  & this work \\
113 & $\nicefrac{1}{2}^+$ & -0.00143(6) & -0.01(4) & \cite{Per11,Fau60}\\
115 & $\nicefrac{1}{2}^+$ & 0.01(2) & 0.05(3)   & this work \\
\noalign{\smallskip}\hline\noalign{\smallskip}
111m & $\nicefrac{11}{2}^-$ & -0.08(4) & -0.10(4) & this work \\
113m & $\nicefrac{11}{2}^-$ & -0.08(2) & -0.08(1) & this work \\
115m & $\nicefrac{11}{2}^-$ & -0.09(4) & -0.09(2) & this work \\
\noalign{\smallskip}\hline\hline\noalign{\smallskip}
Isotope & &  $5s\,5p\,\; ^3\textrm{P}_1$ & & \\
$M$     & $I^\pi$ & $^{111}{\rm \Delta}^{M}~(\%)$ & & Ref.\\
\noalign{\smallskip}\hline\noalign{\smallskip}
107 & $\nicefrac{5}{2}^+$ & -0.0958(8) & & \cite{Per11,Tha63}\\
109 & $\nicefrac{5}{2}^+$ & -0.0912(7) & & \cite{Per11,Tha63}\\
113 & $\nicefrac{1}{2}^+$ & -0.00023(40) & & \cite{Per11,Cha69}\\
\noalign{\smallskip}\hline\noalign{\smallskip}
113m & $\nicefrac{11}{2}^-$ & -0.0773(5) & & \cite{Per11,Cha69}\\
\noalign{\smallskip}\hline
\end{tabular}
\end{table}

\begin{figure}
\resizebox{0.48\textwidth}{!}{%
\includegraphics{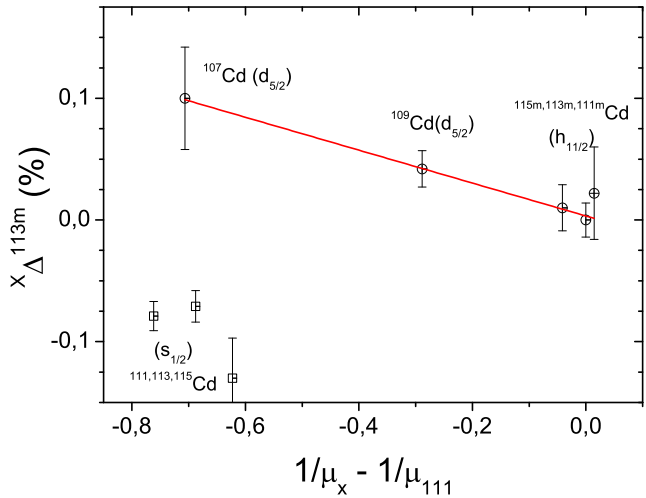}
}
\caption{Moskowitz-Lombardi rule for Cd isotopes: the hfs anomaly for the $5s\,6s\,\; ^3\textrm{S}_1$ level is plotted as a function of the difference of the inverted nuclear moments for nuclear states with $I=\ell + \nicefrac{1}{2}$. A common linear relation is only observed for states with $\ell \neq 0$.}
\label{fig:MoskowitzLombardi}
\end{figure}

Since we have values for the hfs anomaly of several different nuclear states, it is worthwhile to check whether the Moskowitz-Lombardi rule that has been established in the lead region 
\cite{Mos73,Mos82}, works also well 
in the tin region. 
For Cd, all isotopes for which the hyperfine anomaly can be determined are associated with $s_{\nicefrac{1}{2}}$, $d_{\nicefrac{5}{2}}$ and $h_{\nicefrac{11}{2}}$ shell-model states and have nuclear spin $I = \ell+\nicefrac{1}{2}$. To examine whether a similar linear behavior holds for the Cd isotopes, the isotopic hfs anomaly is plotted in Fig.\,\ref{fig:MoskowitzLombardi} as a function of the difference of the inverted nuclear moments for the case of the $5s\,6s\,\; ^3\textrm{S}_1$ level. The distribution looks very similar for $5s\,5p\,\; ^3\textrm{P}_2$ and it is obvious that those isotopes that have a $d_{\nicefrac{5}{2}}$ or $h_{\nicefrac{11}{2}}$ character do indeed exhibit a common linear relation, whereas the isotopes with an $s_{\nicefrac{1}{2}}$ configuration show a distinctly different behavior. The $s$ states belong neither to the $\ell + \nicefrac{1}{2}$ nor to the $\ell - \nicefrac{1}{2}$ branch, because there is no angular momentum $\ell$ to which the spin could couple. Whether the different isotopes with this ground state configuration follow a similar rule with a different slope cannot be decided with the available data. In the case of Hg, there is no similar case since no $s_{\nicefrac{1}{2}}$ isotopes exist in the shell with negative parity filled between $N=82$ and $N=126$. 
The linear fit for the $d_{\nicefrac{5}{2}}$ and $h_{\nicefrac{11}{2}}$ states of Cd produces a downsloping line with $\alpha = -0.135(16) \cdot 10^{-2}\,\mu_N$. This slope is almost an order of magnitude smaller and of opposite sign than in the case of Hg. The corresponding fit for the anomalies extracted from the $5s\,5p \,\; ^3\textrm{P}_2$ level yields a slope of $-0.124(33) \cdot 10^{-2}\,\mu_N$ that agrees well with the one of the $5s\,6s\,\; ^3\textrm{S}_1$ level -- as expected since the differential hfs anomaly almost vanishes. Analyzing the available data from literature in the $5s\,5p\,\; ^3\textrm{P}_1$ level, a much smaller slope of only $\alpha = -0.027(8)  \cdot 10^{-2}\,\mu_N$ is obtained. Finally, our ionic data presented in \cite{Yor13} for the $5s\,\; ^2\textrm{S}_{\nicefrac{1}{2}}$ level result in a slope of $\alpha = -0.065(38) \cdot 10^{-2}\,\mu_N$, being roughly  50\% of the slope for the atomic levels investigated in this work. In all cases the isotopes having nuclear $s$ states must be excluded to observe a linear behavior. It should be noted that the $y$-axis intercept is in all cases compatible with zero within the corresponding uncertainty. 

\subsection{Nuclear magnetic dipole moments}
\label{magnetic_moments}
To determine the nuclear magnetic moments we use Eq.\,(\ref{eq:magnMomentARatio}). The hfs anomaly is negligible if the reference dipole moment is taken from an isotope with the same nuclear spin\,/ parity state, as it has been discussed in the previous section. This is possible for $\nicefrac{5}{2}^+, \nicefrac{1}{2}^+$ and $\nicefrac{11}{2}^-$ isotopes, where we use the reference isotopes $^{109}$Cd, $^{111}$Cd and $^{115\rm{m}}$Cd, respectively. 
The corresponding magnetic moments from literature that were used are included in Table~\ref{tab:moments} and marked with \#. 
For the two isotopes with nuclear spin $I=\nicefrac{3}{2}^+$, $^{121}$Cd and $^{123}$Cd, the magnetic moment of $^{111}$Cd was used as a reference. To estimate the effect of a potential hfs anomaly, the largest value of the determined anomalies ($0.17\,\%$) in Table~\ref{tab:hyperfineanomaly} was assumed and the corresponding effect was added quadratically to the statistical uncertainty of the respective isotope. 
The magnetic moments listed for Cd~I in Table~\ref{tab:moments} are the weighted mean values obtained from the $A$ factors of the $^3\rm{P}_2$ and the $^3\rm{S}_1$ levels, being consistent within uncertainty. They agree with the moments measured by NMR and confirm the measurements on ionic cadmium in \cite{Yor13}. 
\begin{table*}
\centering
\caption{Nuclear magnetic dipole moments and electric quadrupole moments of the cadmium isotopes. For the calculation of the magnetic moments different reference values (marked with $\#$) are used depending on the nuclear spin. The magnetic moments listed for Cd~I are the weighted mean values obtained from the $A$ factors of the $^3\rm{P}_2$ and the $^3\rm{S}_1$ levels. The columns labeled Cd$^+$ show the magnetic moments and the quadrupole moments obtained from collinear laser spectroscopy of ionic cadmium in the $5s\;^2\rm{S}_{\nicefrac{1}{2}} \rightarrow 5p \; ^2\rm{P}_{\nicefrac{3}{2}}$ transition reported in \cite{Yor13}. They are in reasonable agreement with the atomic data. The reference value for the electric quadrupole moments is the quadrupole moment of $^{109}$Cd from \cite{Yor13} (reasons for that are explained in Sect.\,\ref{moments}). The second parentheses for the $Q$-values indicate the systematic uncertainty from the uncertainty of the electric field gradient.}
\label{tab:moments}     
\begin{tabular}{c c l@{}r@{}l r@{}r r@{}r r@{}r@{}r r@{}r@{}l}
\hline\noalign{\smallskip}
 & & & & & \multicolumn{2}{c}{Cd I} & \multicolumn{2}{c}{Cd$^+$ \cite{Yor13}} & \multicolumn{3}{c}{$5s5p\ ^3\rm{P}_2$} & \multicolumn{3}{c}{Cd$^+$ $5p\ ^2\rm{P}_{\nicefrac{3}{2}}$  \cite{Yor13}}\\
\noalign{\smallskip} 
$A$ & $I^\pi$ & \multicolumn{3}{c}{$\mu_{\rm{NMR}}$ ($\mu_{\rm{N}}$)} & \multicolumn{2}{c}{$\mu$ ($\mu_{\rm{N}}$)} & \multicolumn{2}{c}{$\mu$ ($\mu_{\rm{N}}$)} & \multicolumn{3}{c}{$Q_{\rm s}$ (b)} & \multicolumn{3}{c}{$Q_{\rm s}$ (b)}  \\
\noalign{\smallskip}
\hline \hline\noalign{\smallskip}
107 & $\nicefrac{5}{2}^+$ & $-0.6150554$ & (11) & & $-0.6154$ & (3) & $-0.6151$ & (2) & $+0.593$ & (7) & (25) & $+0.601$ & (3) & (24) \\
109 & $\nicefrac{5}{2}^+$ & $-0.8278461$ & (15) & $^\#$ & & & & & & & & $+0.604$ & (1) & (25) $^\#$ \\
111 & $\nicefrac{1}{2}^+$ & $-0.5948861$ & (8) & $^\#$ & & & & & & & & \\
113 & $\nicefrac{1}{2}^+$ & $-0.6223009$ & (9) & & $-0.6224$ & (1)  & $-0.6224$ & (2) & & & & & \\
115 & $\nicefrac{1}{2}^+$ & $-0.6484259$ & (12) & & $-0.6483$ & (2) & $-0.6483$ & (2) & & & \\
117 & $\nicefrac{1}{2}^+$ & & & & $-0.7437$ & (1) & $-0.7436$ & (2) & & & & & \\
119 & $\nicefrac{1}{2}^+$ & & & & $-0.9199$ & (2) & $-0.9201$ & (2) & & & & & \\
121 & $\nicefrac{3}{2}^+$ & & & & $+0.6268$ &(11) & $+0.6269$ & (7) & $-0.288$ & (15) & (12) & $-0.274$ & (7) & (11) \\
123 & $\nicefrac{3}{2}^+$ & & & & $+0.7900$ & (14)& $+0.7896$ & (6) & $+0.063$ & (7)  & (3) & $+0.042$ & (5) & ( 2) \\
\noalign{\smallskip}\hline\noalign{\smallskip}
111m & $\nicefrac{11}{2}^-$ & $-1.1052$ & (2) & & $-1.1052$ & (2) &  $-1.1052$ & (3) & $-0.742$ & (5) & (31) & $-0.747$ & (4) & (30)\\
113m & $\nicefrac{11}{2}^-$ & $-1.0877842$ & (17) & & $-1.0877$ & (2) & $-1.0883$ & (3) &$-0.612$ & (4) & (25) & $-0.612$ & (3) & (25)\\
115m & $\nicefrac{11}{2}^-$ & $-1.0410343$ & (15) & $^\#$ & & & & & $-0.477$ & (4) & (20) & $-0.476$ & (5) & (19)\\
117m & $\nicefrac{11}{2}^-$ & & & & $-0.9969$ & (2) & $-0.9975$ & (4) & $-0.319$ & (2) & (13) & $-0.320$ & (6) & (13)\\
119m & $\nicefrac{11}{2}^-$ & & & & $-0.9636$ & (5) & $-0.9642$ & (3) & $-0.136$ & (3) & (6)\ & $-0.135$ & (3) & ( 5)\\
121m & $\nicefrac{11}{2}^-$ & & & & $-1.0098$ & (3) & $-1.0100$ & (4) & $+0.007$ & (6) & (1) & $+0.009$ & (6) & \\
123m & $\nicefrac{11}{2}^-$ & & & & $-1.0008$ & (5) & $-1.0015$ & (3) & $+0.128$ & (11) & (5) & $+0.135$ & (4) & ( 6)\\
\noalign{\smallskip}\hline
\end{tabular}  
\end{table*}

\subsection{Electric field gradient of the $5s5p\,\; ^3\rm{P}_2$ level in Cd I}
\label{sec:efg}

The quadrupole term of the  hfs formula (\ref{eq:hfsForm}) contains the parameter $B$ which describes the interaction of the nuclear electric quadrupole moment $Q_s$with the electric field gradient $V_{zz}$ produced by the electron cloud at the site of the nucleus, according to
\begin{equation}
   B = e\, Q_s\, V_{zz}\, . 
	\label{eq:Q-int}
\end{equation} 

This electric field gradient (EFG) has been evaluated for the $ 5s5p\ ^3\rm{P}_2 $ excited level of neutral cadmium within the multi-configuration Dirac-Hartree-Fock (MCDHF) theory~\cite{GrantBook2007}. Numerical-grid wave functions have been generated by means of the atomic structure code GRASP~\cite{GRASP2K,grasp3} as self-consistent solutions of the Dirac-Hartree-Fock equations~\cite{Grant:1994}.

\subsubsection{Multi-configuration Dirac-Hartree-Fock theory}
In the MCDHF theory, an atomic level is approximated by a linear combination of configuration state functions (CSF) $ \Phi(\gamma_{k}J) $ of the same symmetry
\begin{equation}                                                                                                 
\label{ASF}
\Psi_\alpha(J) \:=\: \sum_{k}^{\rm NCF} c_{k} (\alpha) \: \Phi(\gamma_{k}J),
\end{equation}
where ${\rm NCF}$  refers to the number of CSFs and $\{ c_k (\alpha) \}$ to the representation of the atomic level in the given basis. Moreover, the sets $ \gamma_{k} $ describe all quantum numbers that are required to
distinguish the basis states uniquely. In the GRASP code, the CSFs $\Phi(\gamma_{k}J)$ are constructed as antisymmetrized products of a \textit{common} set of orthonormal orbitals, and are optimized together on the basis of the Dirac-Coulomb Hamiltonian
\begin{equation}
\label{Dirac-Coulomb-Hamiltonian}
H_{DC} = \sum_{i} \left[ c {\bm{ \alpha }}_i \cdot
                    {\bm{ p }}_i
         + (\beta_i -1)c^2 + V(r_i) \right]
         + \sum_{i>j} 1/r_{ij} \,.
\end{equation}
Further relativistic effects due to the Breit interaction could, in principle, be added to the representation $\{ c_k(\alpha) \}$ by diagonalizing the Dirac-Coulomb-Breit Hamiltonian \linebreak[4] matrix but they are known to have little effect upon the EFG and hyperfine parameters of light- and medium-Z elements~\cite{Paduch:Xe-Q:2000,Derevianko:2001,Bieron:Br_I-Q:2001}. For the $ 5s5p\ ^3\rm{P}_2  $ level of neutral cadmium only CSFs of even parity and total angular momentum $J=2$ need to be taken into account in the wave function expansion. 

\subsubsection{Wave function generation}
\label{wavefunctiongeneration} 
The computational methodology of generating wave functions for hyperfine calculations have been explained elsewhere~\cite{Bieron:Li:1996,Bieron:Au:2009}. We shall describe here only the basics of the models which were applied in the three series of computations. All three computational models share a common set of spectroscopic (occupied) orbitals. All spectroscopic orbitals $1s,\, 2s,\, ...,\, 4d,\, 5s,\, 5p$ of the $ 5s5p\ ^3 \rm{P}_2 $ level of neutral cadmium were obtained in the Dirac-Hartree-Fock approximation and were frozen in all further steps of the computations. Properly antisymmetrized linear combination of products of the set of spectroscopic orbitals formed the zero-order representation (often referred to as a reference configuration state function) of the $ 5s5p\ ^3 \rm{P}_2 $ atomic level.

Then two sets of virtual orbitals were generated in a series of steps in order to incorporate electron-electron correlations at increasing level of (computational) complexity and to monitor their effect upon the level energy and the EFG of the $ 5s5p\ ^3\rm{P}_2 $ level. Our computational models differ in the way the virtual orbitals are generated, as well as in the sets of CSFs in the wave function expansion (\ref{ASF}); see~Refs.~\cite{Bieron:Sc:1997,Bieron:Au:2009} for further information. Below, we briefly refer to these models as \textit{single substitutions} ({\sl S}) and \textit{single with restricted double substitutions} ({\sl SrrD} and {\sl SrD});
 in some more detail, these models include:
\begin{description}[labelindent=1cm, align=right, leftmargin=*]
\item[{\sl S}]
Single substitutions from the valence and core orbitals to eight layers of virtual orbitals. Within first-order perturbation theory, only single substitutions contribute to the hyperfine energy~\cite{Lindgren84,Wybourne65-148}. In the MCDF method, in contrast, double substitutions from the (sub-) valence shells often dominate energetically over correlation corrections and, hence, the virtual orbitals obtained in correlated variational calculations are determined predominantly through the effects of double substitutions upon the total level energy. For this reason, however, it is important that the virtual space is constructed from orbitals optimized for single substitutions~\cite{Engels93}.
\item[{\sl SrrD}] 
Single and restricted double substitutions from the valence and core orbitals to seven layers of virtual orbitals. Double substitutions were restricted in the sense that at most one electron is promoted from core orbitals, while the other (or both) is (are) promoted from the valence shells $5s5p$. The size of the multi-configuration expansions is further restricted in this model by eliminating all CSF which do not  interact with the reference CSF. Calculations based on an {\sl SrrD} computational mo\-del have been found optimal for {\sl ab initio} calculations of hyperfine  structures in heavy atoms~\cite{Bieron:Au:2009}.
\item[{\sl SrD}] 
The orbital sets as generated in the step {\sl SrrD}  above were utilized also for performing con\-fig\-ura\-tion-interaction calculations (i.e.~with all orbitals frozen), where the CSF reductions described in the step {\sl SrrD} above were lifted. For such expansions, the {\sl SrD} approximation can be considered as a  consistency check on the wave functions obtained in the {\sl SrrD} model.
\end{description}
For the two models {\sl S} and {\sl SrrD}, virtual orbitals were generated stepwise in terms of \textit{layers}~\cite{Fritzsche:02,Bieron:Au:2009}, until saturation of the calculated electric field gradient (EFG) was observed. Eight layers of virtual orbitals were generated in the approximation {\sl S} by opening eventually all core orbitals, down to $1s$, for substitutions. For the {\sl SrrD} and {\sl SrD} models, a total of seven layers of correlation orbitals were generated.

\begin{figure}
\resizebox{0.5\textwidth}{!}{%
\includegraphics{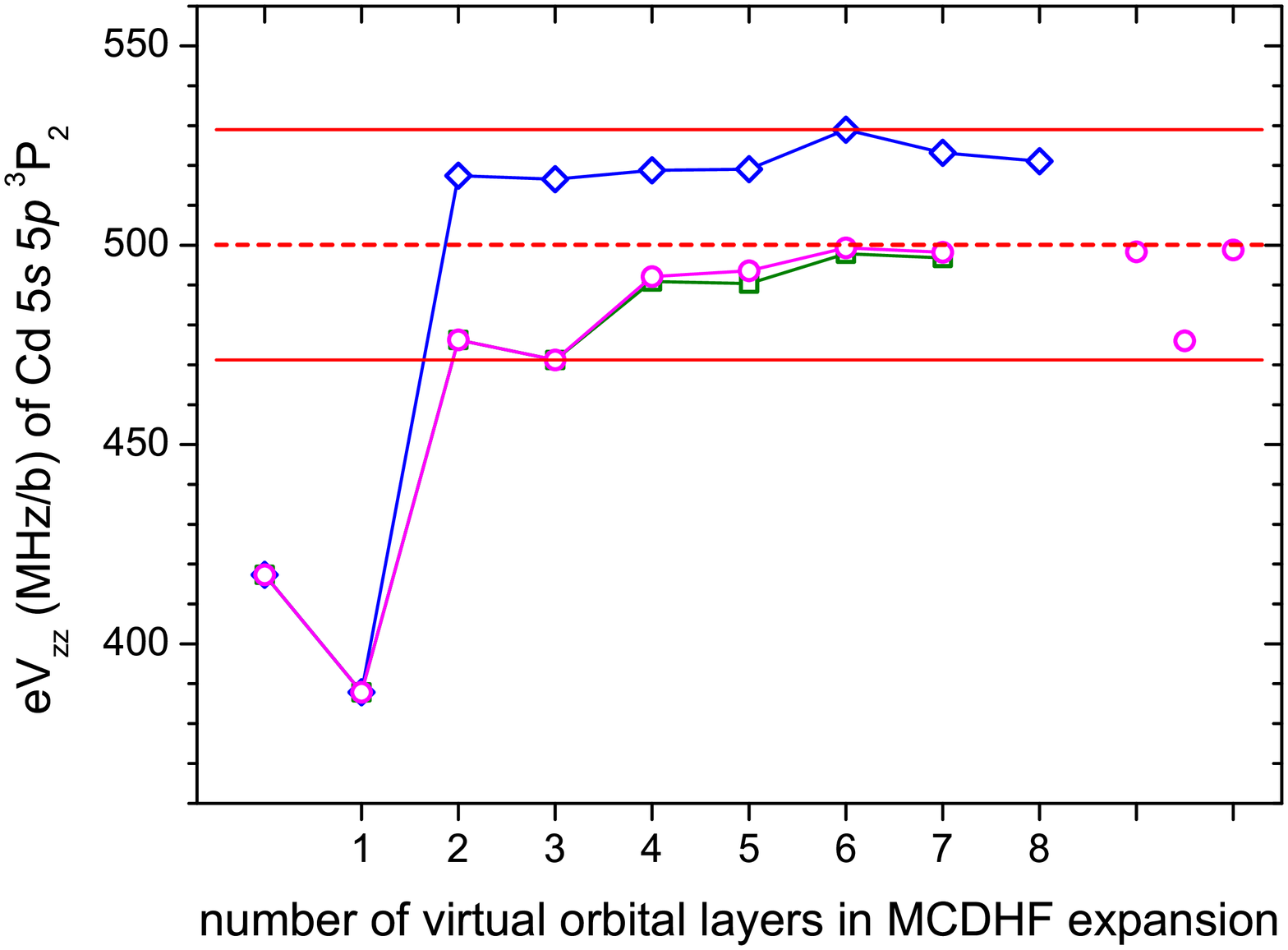}}
\resizebox{0.5\textwidth}{!}{%
\includegraphics{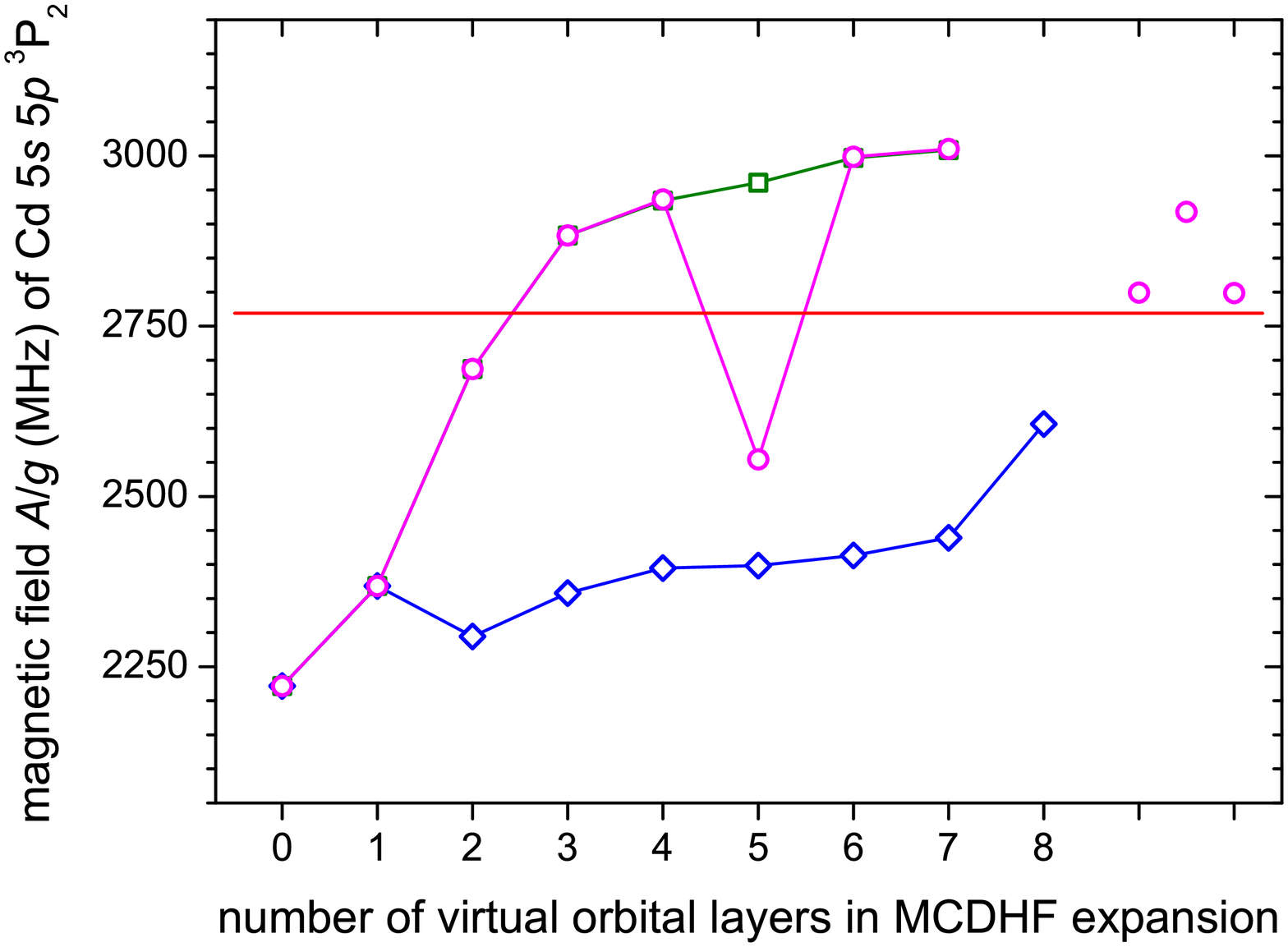}}
\caption{(Color online) Electric field gradient $eV_{zz}$ (top) and magnetic field ($A/g$, bottom) for the low-lying excited $5s5p\ ^3\rm{P}_{2}$ level of neutral cadmium and for the three computational models as described in the text, i.e.~by including single substitutions {\sl S} (blue diamonds); single with restricted double substitutions {\sl SrrD} (olive squares); and {\sl SrD}  expansions (magenta circles). The integers on the abscissa axis represent the number of (correlation) layers of virtual orbitals that are taken into account in the expansion (\ref{ASF}). The three (magenta) circles to the right side of the figure display configuration-interaction calculations which include full double and triple substitutions. All lines are drawn only for the guidance of the eyes. The two (red) horizontal lines without symbols in the top graph denote the estimated uncertainty of the final computational result of $eV_{zz}$, which is indicated by the dashed line. In the bottom graph, the (red) horizontal line represents the experimental value of $A/g$. See text for further details.}
\label{efg-figure}
\end{figure}

In {\sl ab initio} calculations of hfs, double and triple substitutions are known to cancel each other to a large degree~\cite{Engels93}, and their combined effect upon the calculated EFG values is typically a few percent or even less. This was confirmed by a number of earlier calculations on other medium and heavy elements~\cite{Bieron:Sc:1997,Bieron:Hg:2005,Bieron:Au:2009}. Such unrestricted double and triple 
substitutions were included also in our recent calculations for the excited $ 5p\ ^2\rm{P}_{\nicefrac{3}{2}} $ level of $ ^{109}$Cd$^+$ ion~\cite{Yor13}, in which these double and triple substitutions contributed individually less than 3 percent each, and below 1 percent when considered together (i.e.~well below the estimated 4 percent accuracy of the final EFG value in Ref.~\cite{Yor13}). In the present work, we therefore performed only three large-scale configuration interaction calculations, in which we compared the effects of unrestricted double against unrestricted triple substitutions. These computations were carried out for (i)  substitutions from $4s\-4p\-4d\-5s\-5p$ occupied orbitals to one layer of virtual orbitals; and (ii) for substitutions from the $3s\-3p\-3d\-4s\-4p\-4d\-5s\-5p$ occupied orbitals to one layer. The results of these three configuration-interaction calculations are
represented by the three unconnected (magenta) circles on the right side of figure~\ref{efg-figure}. They clearly indicate that the corrections arising from the unrestricted double and triple substitutions are smaller than the estimated 4 percent accuracy of the final EFG value.
\newline The wave functions have been generated with the nucleus modelled as a sphere, and a two-parameter Fermi distribution~\cite{grasp89} was employed to approximate the radial dependence of the nuclear charge density. All other nuclear electromagnetic moments were assumed to be point-like, i.e.~the magnetization distribution inside the nucleus (the Bohr-Weisskopf effect~\cite{BohrWeisskopf1950}), and still higher nuclear electromagnetic moments, as well as Breit and QED corrections~\cite{graspMcKenzie1980}, were all neglected, since they are expected to be negligible at the current level of accuracy. These effects were found to be small even for an element as heavy as radium~\cite{Bieron:Ra-Q:2005,Bieron:Ra-hfs:2005}. All further details of the computations, including a method to take advantage of the computational balance between the hyperfine $A$ and $B$ 
coefficients in the evaluation of EFG, will be described elsewhere.

\subsubsection{EFG theory}
The quadrupole moment $Q$ of an isotope is related to the (electric-quadrupole) hyperfine constant $B$ through the equation
\begin{equation}
\label{BeQT}
 B(J)  = 2 e \, Q \, \left< JJ | T^{(2)} | JJ \right> \, ,
\end{equation}
with $e$ being the charge of the electron, and where the (second-rank) operator $ T^{(2)} $ acts upon the electronic coordinates of the wave functions. It is this operator in the theory of  hfs \cite{Lindgren84,Wybourne65-148} that describes the electric field gradient, EFG, at the site of the nucleus
\begin{equation}
\label{T2Vzz}
\left< JJ | T^{(2)} | JJ \right> = \frac{1}{2} \left<	 \frac{\partial^2 V}{\partial z^2} \right>\,.
\end{equation}

\subsubsection{EFG results}
Figure~\ref{efg-figure} displays the calculated values of the EFG for the $5s5p\ ^3\rm{P}_{2}$ excited level of neutral cadmium for different computational models at various degrees of complexity. Results are shown for 
the three (computational) models as explained above, i.e.~by including: single substitutions {\sl S} (blue diamonds), single with restricted double substitutions {\sl SrrD} (olive squares), as well as for the {\sl SrD} 
configuration-interaction calculations (magenta circles).

All lines in this graph are drawn only for the guidance of the eyes. The computational complexity is represented by the number of (correlation) layers on the abscissa axis that are taken into account in the expansion (\ref{ASF}) of the corresponding wave functions. Up to 3868 CSFs were included in the expansions for single substitutions into eight correlations layers, and up to 216643 CSFs in the {\sl SrD} model in the final stages  of the computations. The uncorrelated Dirac-Hartree-Fock value ($n=0$) is also shown but is not considered for estimating the uncertainty of the computations. Our largest expansions yield the value 
$eV_{zz} = 500 \pm 30 $\, MHz/b, where the uncertainty was assumed to cover all values for the layers of virtual orbitals with $n = 2,...,8$. This conservative estimate of the `uncertainty' due to missing correlations is shown by the two horizontal lines (without symbols) in Fig.~\ref{efg-figure}.

While the final value $eV_{zz} = 500.108$ MHz/b from the largest computation almost coincides (perhaps accidentally) with the $n = 7$ points of the {\sl SrrD} and {\sl SrD} approaches in Fig.~\ref{efg-figure}, single substitution model {\sl S} yields the final value close to the upper straight (uncertainty) line. As mentioned above, the two models {\sl SrrD} and {\sl SrD} are considered to be suitable especially for {\sl ab initio} calculations of hyperfine structures. Therefore, it appears well justified to assume the last points of the {\sl SrrD} and {\sl SrD} curves as final value, and the deviations of all three curves together as a measure for the uncertainty of the computations.

As a test of their reliability, the same wave functions were used to calculate the magnetic field produced by the electrons at the site of the nucleus, and subsequently to compare with the measured A-factors and magnetic moments.
The bottom graph in Fig.~\ref{efg-figure} displays the calculated values of the magnetic field $A/g$ for the $5s5p\ ^3\rm{P}_{2}$ excited level of neutral cadmium, obtained from the same computational models as described in Sec.~\ref{wavefunctiongeneration} above and in the caption of Fig.~\ref{efg-figure}. The horizontal (red online) straight line represents the experimental value of $A/g$. These graphs support the conclusions drawn in the preceeding paragraphs, concerning the uncertainty of the computations. The graphs clearly indicate that the calculations of the magnetic field have not produced a converged value of $A/g$, but $A/g$ is not the objective of this work.

From the above analyses, we conclude a  (theoretical) value $eV_{zz} = 500 \pm 30 $\, MHz/b, which can be utilized in deriving the quadrupole moments from the measurements of the hyperfine $B$ parameters.

\subsection{Electric quadrupole moments}
\label{moments}
The nuclear quadrupole moments can now be directly extracted from the determined spectroscopic $B$ factors using the calculated EFG value of the $ 5s5p\ ^3 \rm{P}_2 $ level. For $^{109}$Cd, we obtain $Q_{\rm s}(^{109}\mathrm{Cd})=592(1)(36)$\,b. The first parantheses represent the statistical measuring uncertainty and the second ones the EFG-related systematic uncertainty. The latter is about 50\% larger than in the value $Q_{\rm s}(^{109}\mathrm{Cd})=0.604(1)(25)$\,b, reported in \cite{Yor13} from the $5p\, ^2 \rm{P}_{\nicefrac{3}{2}}$ level of the Cd$^+$ ion and based on $eV_{zz} = 666(27)$\,MHz/b. This EFG for the Cd$^+$ level was calculated in the same way as described in the previous section. The quadrupole moment extracted directly from the atomic level is only 2\% smaller than the result obtained for the ion, a difference that is insignificant at the level of accuracy of the EFG values. Hence, the results obtained in two different ways clearly demonstrate that the EFG calculations for the atomic and the ionic systems are compatible and provide consistent data. In order to provide a coherent set of quadrupole moments from both measurements with the smallest possible uncertainty, we prefer to base all spectroscopic quadrupole moments on the ionic result by using $Q_\mathrm{s}(^{109}\mathrm{Cd})=0.604(1)(25)$\,b as reference value and employing the relation 
\begin{eqnarray}
   Q_\mathrm{s}\left(^{A}\mathrm{Cd}\right) = \frac{B\left(^{A}\mathrm{Cd}\right)}
	{B\left(^{109}\mathrm{Cd}\right)}\cdot Q_\mathrm{s}\left(^{109}\mathrm{Cd}\right)  \nonumber
\end{eqnarray}
for the other isotopes. The results are listed in Table~\ref{tab:moments}. 
The most intriguing observation is the linear increase of the electric quadrupole moments of the $\nicefrac{11}{2}^-$--isomers that has been extensively discussed in \cite{Yor13}. The values reported here are in excellent agreement with the ionic results. It should be noted that using the directly calculated EFG value for the atomic level would simply result in a reduction of all quadrupole moments by 2\%. 

The literature values \cite{Sto05} of the quadru\-pole moments of $^{105}$Cd, $^{107}$Cd, $^{109}$Cd, $^{111m}$Cd, $^{113m}$Cd and $^{105}$Cd are more than 10 \% larger than the present results. They are all based on the hfs of the $5s\,5p\;\, ^3\mathrm{P}_1$ level of Cd I and a semi-empirical evaluation of the EFG using information from the magnetic hyperfine interaction and the fine structure splitting \cite{Lau69} and neglecting Sternheimer-type shielding correction. A systematic uncertainty from this procedure was estimated to be 10 \%, which is in accordance with the deviation from our results.

Finally, the isotope shifts with respect to $^{114}$Cd and the isomer shifts between the isomeric and the respective ground state are extracted from the centers of gravity obtained from the hfs fits. These will be discussed in a separate publication combined with similar information from the ionic $5s\,^2\mathrm{S}_{\nicefrac{1}{2}} \rightarrow 5p\,^2\mathrm{P}_{\nicefrac{3}{2}}$ transition.

\section{Summary}
\label{summary}
Nuclear spins and hfs parameters of $^{107-123}$Cd and the isomers $^{111m-123m}$Cd were determined by collinear laser spectroscopy. The nuclear spin of $^{119}$Cd was unambigiously determined as $I=\nicefrac{1}{2}$. Nuclear magnetic dipole moments and electric quadrupole moments were extracted using reference values and results of EFG calculations, respectively. Hyperfine structure anomalies and differential hfs anomalies 
were investigated and the validity of the Moskowitz-Lombardi rule was discussed. While the anomalies of nuclear $d_{\nicefrac{5}{2}}$ and $h_{\nicefrac{11}{2}}$ states show a linear behavior comparable to that observed in the Hg isotopes, the isotopes with an $s_{\nicefrac{1}{2}}$ ground state having $\ell=0$ clearly deviate from the rule. 
The magnetic moments and quadrupole moments are in excellent agreement with values measured in the ionic system \cite{Yor13}. 

\section{Acknowledgement}
\label{acknowledgement}
\begin{acknowledgement}
This work has been supported by the Max-Planck Society, the German Federal Ministry for Education and Research under Contract No. 05P12RDCIC,  the Helmholtz International Center for FAIR (HIC for FAIR) within the LOEWE program by the State of Hesse, the Belgian IAP Project No. P7/12, the FWO-Vlaanderen, and the European Union seventh framework through ENSAR under Contract No. 262010. 
N. F. received support through GRK Symmetry Breaking (DFG/GRK 1581). We thank the ISOLDE technical group for their professional assistance.
The theory part of this work was supported by the Polish Ministry of Science and Higher Education (MNiSW) in the framework of the project No.~N~N202~014140 awarded for the years 2011-2014.
The large-scale calculations were carried out with the supercomputer Deszno purchased thanks to the financial support of the European Regional Development Fund in the framework of the Polish Innovation Economy Operational Program (contract no. POIG.02.01.00-12-023/08).

\end{acknowledgement}

\bibliography{biblio}
\bibliographystyle{phreport}

\end{document}